\documentclass[aps,prl,twocolumn,superscriptaddress,amsmath,amssymb]{revtex4-1}

\usepackage{graphicx}
\usepackage{color}
\usepackage{float}
\usepackage{hyperref}
\begin{document}

\title{Beyond Quality and Quantity: Contact Distribution Encodes Frictional Strength}

\author{Sam Dillavou}
\affiliation{John A Paulson School of Engineering and Applied Sciences, Harvard University, Cambridge, MA 02138, USA}
\affiliation{Department of Physics, Harvard University, Cambridge, MA 02138, USA}
\affiliation{Department of Physics and Astronomy, University of Pennyslvania, PA 19104,USA}
\author{Yohai Bar-Sinai}
\affiliation{John A Paulson School of Engineering and Applied Sciences, Harvard University, Cambridge, MA 02138, USA}
\affiliation{The Raymond and Beverly Sackler School of Physics and Astronomy, Tel Aviv University, Tel Aviv 69978, Israel}
\author{Michael P. Brenner}
\affiliation{John A Paulson School of Engineering and Applied Sciences, Harvard University, Cambridge, MA 02138, USA}
\affiliation{Google Research, Mountain View, CA 94043, USA}
\author{Shmuel M. Rubinstein}
\affiliation{John A Paulson School of Engineering and Applied Sciences, Harvard University, Cambridge, MA 02138, USA}
\affiliation{Racah Institute of Physics, The Hebrew University of Jerusalem, Jerusalem, 91904, Israel}

\date{\today}
\begin{abstract}
Classically, the quantity of contact area $A_R$ between two bodies is considered a proxy for the force of friction. However, bond density across the interface - quality of contact - is also relevant, and contemporary debate often centers around the relative importance of these two factors. In this work, we demonstrate that a third factor, often overlooked, plays a significant role in static frictional strength: the distribution of contact. We perform static friction measurements, $\mu$, on three pairs of solid blocks while imaging the contact plane. By using linear regression on hundreds of image-$\mu$ pairs, we are able to predict future friction measurements with 3 to 7 times better accuracy than existing benchmarks, including total quantity of contact area. Our model has no access to quality of contact, and we therefore conclude that a large portion of the interfacial state is encoded in the spatial distribution of contact, rather than its quality or quantity. 
\end{abstract}


\maketitle

\newcommand{\appropto}{\mathrel{\vcenter{
  \offinterlineskip\halign{\hfil$##$\cr
    \propto\cr\noalign{\kern2pt}\sim\cr\noalign{\kern-2pt}}}}}

Static friction, the force required to initiate sliding between two solid bodies, is an illusive quantity that is famously difficult to predict precisely. This reflects the fact that this force is a single scalar which is the outcome of a complex spatio-temporal process of slip nucleation across a typically heterogeneous interface, and as a result depends on a large variety of factors, both controlled \cite{Baumberger:2006bq, Li:2011gf, Bocquet:1998wt, MasaoNakatani:1996wu, Berthoud:1999ha,Bureau:2002br, Karner:2000aa,Karner:1998ik,BenDavid:2011di,Dillavou:2018in} and uncontrolled (such as wear) \cite{Dillavou:2018in,BenDavid:2011di}. Even in well-designed, rigorous laboratory experiments static friction can vary significantly and unpredictably between successive measurements using the same two bodies \cite{BenDavid:2011di,Pitenis:2014eb}. This stochasticity largely stems from one inconvenient truth about frictional interfaces: even using the same bulk solids, a new system is formed after each slide. Each such interfacial system contains the ensemble of contact points between two rough bodies, which typically covers a small fraction of the interface due to surface roughness.  The frictional strength is classically considered a linear function of the total real contact area of an interface $A_R$, as the two quantities generally evolve in tandem \cite{BowdenTabor,Archard:1957up,Greenwood:1966boa,Persson:2001kz,Dieterich:1994ux, Berthoud:1999ha, Baumberger:2006bq}. 

Several exceptions to the proportionality between $A_R$ and $\mu$, the static coefficient of friction, were demonstrated recently in the context of frictional aging (strengthening over time) \cite{Weber:2019aa,Li:2011gf}. Typically, these works conclude that time-dependent quality of contact - the density of chemical bonding across the interface - explains the discrepancy \cite{Liu:2012hua,Tian:2017cn}; that is, frictional strength can still be thought of as a function of integrated contact area, albeit appropriately weighted by contact quality. This framework is appealing, as it reduces the relevant state of the entire contact ensemble to one number, consistent with the state-of-the-art predictive model for friction, known as Rate and State friction laws \cite{Dieterich:1979vq, Rice:1983aa, Ruina:1983hh}. However, a growing body of evidence suggests that the relevant interfacial state is in fact more complex than a single number can describe \cite{Karner:1998ik,BenDavid:2011di,Dillavou:2018in,Dillavou:2020cs, Pilvelait:2020ki}. Rate and State friction laws are therefore a reasonable but crude approximation of static frictional strength and its details remain the subject of continual debate~\cite{Bhattacharya:2017fy}, while the degree of complexity required to model frictional strength is still an open question.

Predicting a single number, such as $\mu$, from a complex data set is a canonical problem in data science. Noteworthy progress has recently been made in predicting laboratory or real earthquakes by utilizing machine learning methods like convolutional neural networks or boosted decision trees, \cite{RouetLeduc:2017fk, Hulbert:2018jw,Perol:2018cu,Corbi:2019gz}. In closely related works, similar methods were used to predict mechanical failure of rocks \cite{McBeck:2020gd} and amorphous solids \cite{Cubuk:2015cd}. Most work utilizes signals that do not provide direct measurement of the internal interfacial state, meaning that even successful predictions are difficult to interrogate. Some prediction work has been done using direct measurements from bi-material model faults \cite{Corbi:2019gz}, but with equally complex algorithms, and it is unclear if and how these results may apply to single material, multi-contact interfaces. Together, these exciting results indicate that friction is more predictable than previously thought, suggesting the possibility of predictive models using direct measurements of the interface in a straightforward and transparent manner.

\begin{figure} 
\includegraphics[width = .48\textwidth]{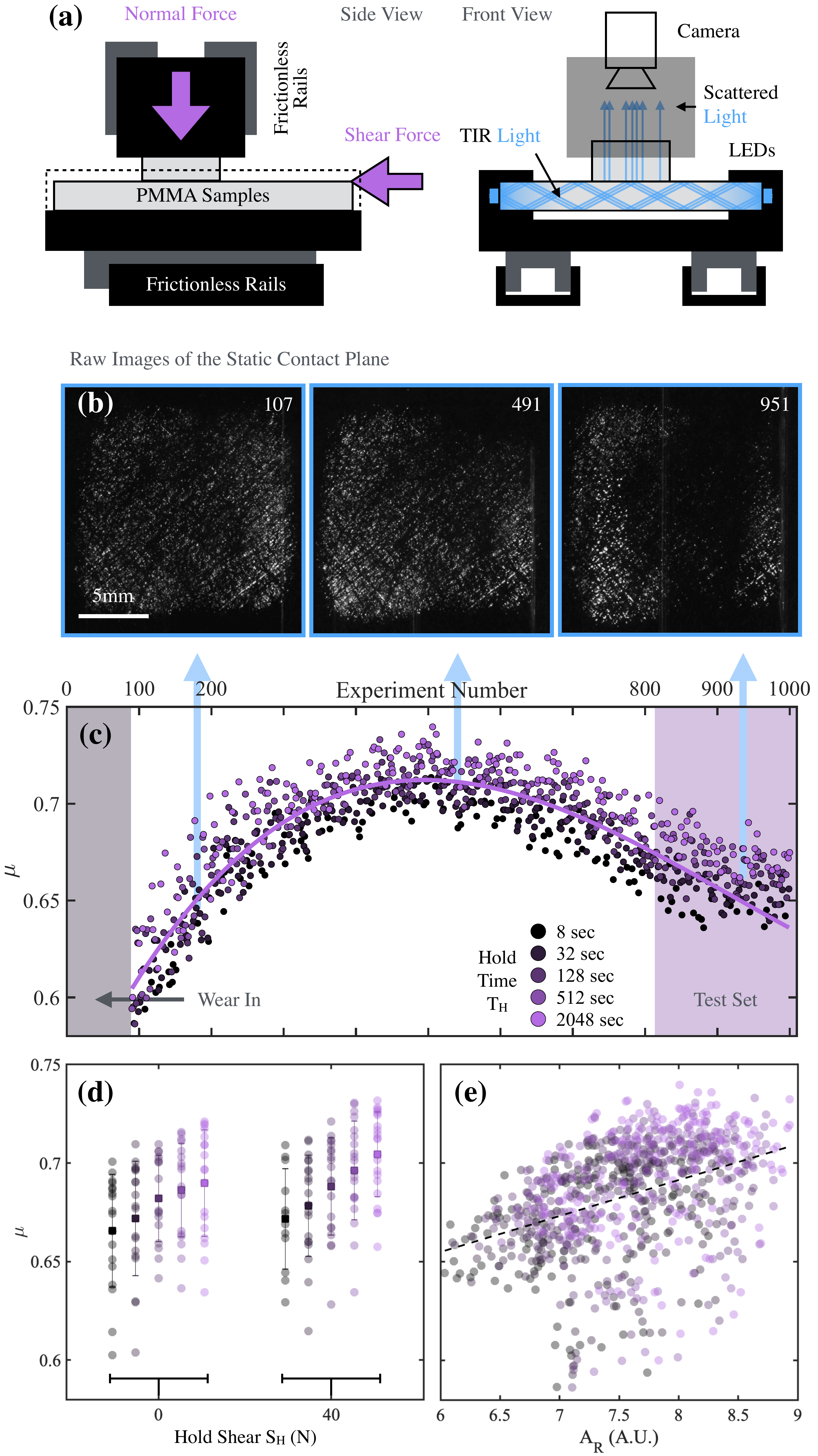}
\caption{\textbf{Experimental Setup and Benchmark Parameters} a) Left: Schematic of the biaxial compression/translation stage. Right: Embedded optical setup to image contact ensemble. b) Three typical images of interfacial contact for after a few dozen, a few hundred, and nearly one thousand experiments, respectively. All data in (b-e) is for block pair 1. c) $\mu$ vs experiment number. Colors in (c-e) indicate experimental parameter hold time $T_H$. Final 18\% of data, to be used as the test set for predictions, is highlighted in purple on the right. d) $\mu$ for the highest (40N) and lowest (0N) hold shears $S_H$ for block pair 1, separated and color-coded by hold time. Circles are individual experiments, squares are means and error bars are standard deviations for unique ($S_H$,$T_H$) pairs. e) $\mu$ as a function of real area of contact $A_R$ (sum of image intensities). Dotted line is a linear fit to all data.
\label{1}}
\end{figure} 





Here we use linear regression to predict the static friction coefficient of a multi-contact interface undergoing frictional aging using spatially resolved images of its real area of contact and no other inputs. This method is 3 to 7 times more accurate than the benchmark methods of prediction using using the total area of contact and experimental parameters. Our results indicate that frictional strength is encoded in the spatial distribution of the real area of contact.

The biaxial compression and translation stage used to measure the friction coefficient is described in detail in a previous work \cite{Dillavou:2018in}, and shown schematically in Fig \ref{1}(a). Experiments are performed separately on three pairs of laser-cut PMMA (poly methyl-methacrylate) blocks with 1 - 2.5 cm$^2$ of nominal contact area. The bottom samples are original, extruded PMMA (11nm RMS), approximately 60x100x4mm, which are directly contacted by the horizontal force sensor. The top samples are lapped with 1000 grit polishing paper ($\sim$800nm RMS), and are the main source of variance between interfacial systems.  While the samples are in contact, the interface is imaged using a total internal reflection (TIR) technique: single-wavelength (473nm) light is injected into the bottom sample where it remains trapped though TIR, except at points of actual contact with the top sample. As a result, when imaged from above, the brightness of the interface corresponds to points of real contact, as shown for three examples in Fig.~\ref{1}(b). The camera position is fixed in relation to the top (smaller, rougher) block, and thus images in subsequent experiments contain common features. Images have a resolution of approximately 1 pixel per 10 $\mu$m, the same order of magnitude as one contact point.

Static friction measurements are taken via the standard Slide-Hold-Slide (SHS) protocol: Under constant normal load, $F_N = 90N$, samples are slid at 0.33 m/s to create a new contact ensemble. The interface is then held static for hold time $T_H$ sec under constant hold shear $S_H$. At the last moment the interface is held static, the image of the contact plane is taken. Subsequently the horizontal motor switches to position-control and loads the interface at a rate of 0.33 mm/s ($\sim$33 N/s) until the initiation of slip, accompanied by a sharp drop in the measured shear force. We define $\mu$ as the highest shear force prior to slip, or the `static peak', divided by the normal load. 

Over the course of hundreds of experiments, repeated sliding slowly wears the surfaces of our samples. This effect manifests in changing of the contact ensemble, and generates a slow, and non-monotonic, drift of the friction coefficient, as shown in Figs \ref{1}(b) and (c). This effect is most rapid with a fresh sample, and thus the first several dozen experiments are discarded from our data set (``run-in''). Regardless, to avoid conflating the effect of wear with the effects of changing hold time $T_H$ or hold shear $S_H$, the experiments are ordered such that every possible combination of experimental variables is performed once in a random order, then again in a different random order and so on. At least five unique values of $S_H$ and of $T_H$ are used for each block, see \footnote{Block Pair 1: $S_H=$ \{0, 5, 10, 15, 20, 25, 30, 35, 40\}, $T_H=$ \{8, 32, 128, 512, 2048\}. Block Pair 2: $S_H=$ \{0, 5, 10, 15, 20, 25, 30, 35, 40\}, $T_H=$ \{16, 32, 64, 128, 256, 512\}. Block Pair 3: $S_H=$ \{-30, -15, 0, 15, 30\}, $T_H=$ \{8, 16, 64, 256, 1024\}.} for details. 

Static frictional strength $\mu$ has a systematic but noisy dependence on several factors in our data. For example, it is well established that static friction `ages', that is, it is correlated  with both the logarithm of the hold time $\log(t)$~\cite{Dieterich:1979vq, Ruina:1983hh, Baumberger:2006bq}, and this logarithmic rate is dependent on hold shear $S_H$~\cite{Heslot:1994gd,Dillavou:2020cs}. Our data shows this expected dependence, as demonstrated in Fig \ref{1}(c) and (d), and also displays a weak dependence on real area of contact $A_R$, as shown in Fig \ref{1}(e). Note that the classical relationships are present in aggregate; $A_R$, $\log(t)$, and $\mu$ are all positively correlated. However, these correlations are swamped by noise, and do relatively little to predict frictional strength for an individual experiment. In previous works with this experimental setup \cite{Dillavou:2018in,Dillavou:2020cs}, wear was treated as a confounding variable, and its resulting slow non-monotonic trend was subtracted from $\mu$ to highlight the effect of experimental parameters. This technique is discussed later in this report as another benchmark bested by our method, however a true prediction of $\mu$ should not involve any such modification of the data. With or without such de-trending, there is a large variance in $\mu$ that is not accounted for by experimental parameters and $A_R$, but, as we show, is in large part predictable from the contact distribution.

\begin{figure} 
\includegraphics[width = .38\textwidth]{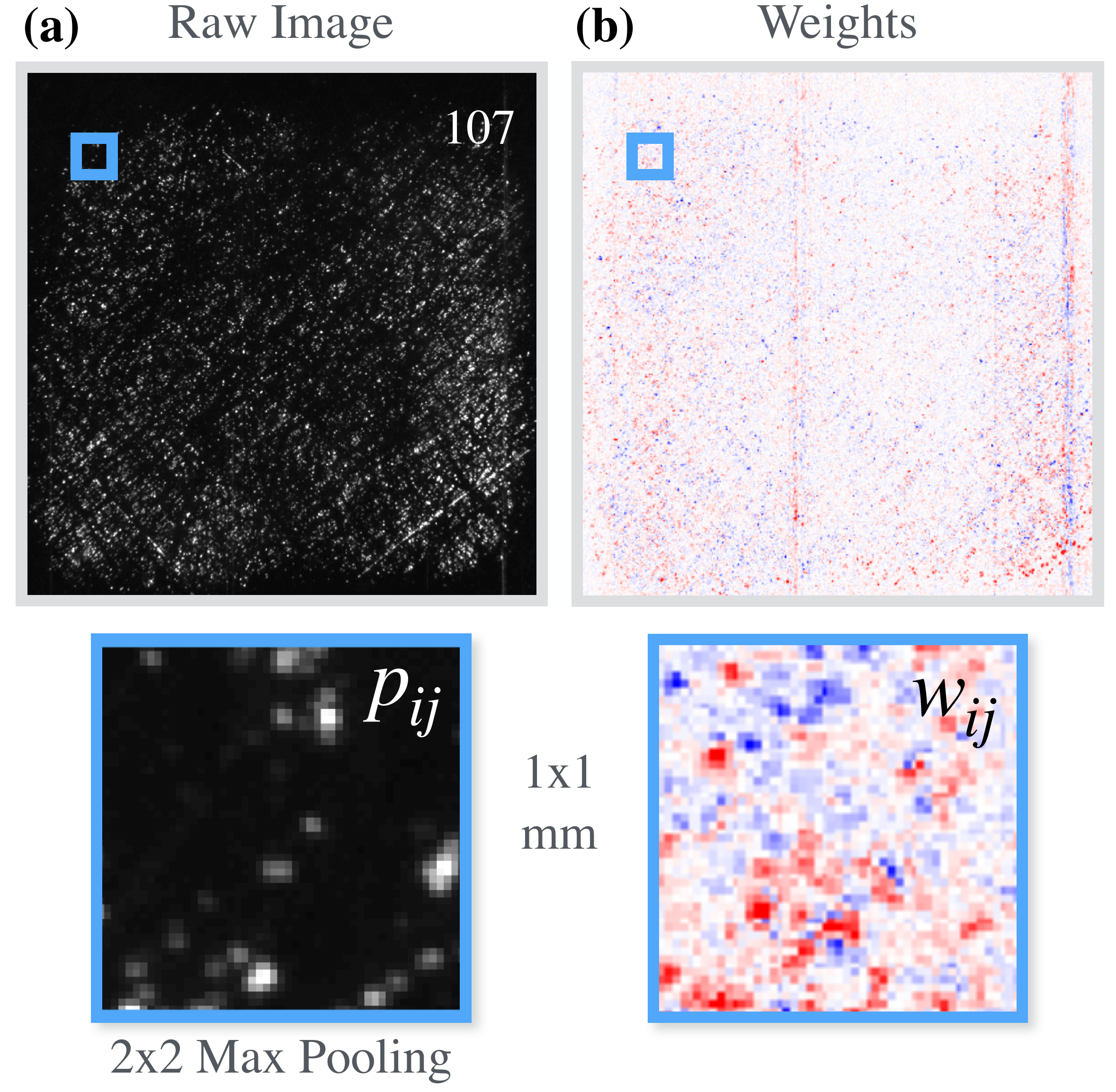}
\caption{\textbf{Image Processing and Weights} a) Example raw image from block pair 1, and 1mm$^2$ section (blue outline) after 2x2 max pooling. b) Visualization of the resulting $w_{ij}$ with a zoom in on 1mm$^2$ section from (a). Red is positive, white is zero, and blue is negative.  
\label{2}}
\end{figure} 

We now turn to the distribution of the contact ensemble to explain this variance. Like most physical (non-digital) systems, our data collection is limited by real-world constraints \cite{Hoffmann:2019by}. Block pairs can be used for only a few hundred to one thousand experiments before they are worn beyond use. Since each image contains millions of pixels (``features''), but each block can only provide $\sim1,000$ examples, the problem is massively under-constrained, and we reduce the complexity of our model slightly by square-kernel max pooling by a factor of 4. This reduction speeds computation, and smooths out small-scale details, as shown in Fig \ref{2}(a). 

Friction predictions $\hat \mu$ are constructed using linear regression of gray-scale pixel intensities $p_{ij}$ of these reduced images. Explicitly
\begin{equation}
\hat \mu(p_{ij}) = C + \sum_{ij} p_{ij}w_{ij}
\label{prediction}
\end{equation}
where C and $w_{ij}$ are fitting parameters (weights) that are constant for each block pair. These are found by standard Ridge regression~\cite{Bishop}, i.e.~a regularized minimization of the prediction error,
\begin{equation}
\operatornamewithlimits{argmin}_{w_{ij},C} \left\{ \sum_n \left(\mu^{(n)} - 
\hat{\mu}\left(p_{ij}^{(n)}\right)\right)^2 + \alpha\sum_{ij} w_{ij}^2 \right\} \ ,
\label{cost}
\end{equation}
where $\mu^n$ and $p_{ij}^{(n)}$ are the static friction coefficient and the interfacial image of the $n$-th experiment. $\alpha$ is a hyper-parameter that discourages over-fitting. For each block, the first 82\% of experiments are used for training and cross-validation (that is, finding the optimal $\alpha$ through leave-one-out cross validation~\cite{Bishop}). All metrics reported below are evaluated on the last 18\% of the data, which were not used during training. Our model produces $w_{ij}$ that have the size, shape, and granularity of the reduced images, as shown for typical values in Fig. \ref{2}(b) for block pair 1. These values are therefore not transferable from pair to pair, as they relate to specific asperities of a single pair. As our interfaces experience irreversible evolution through wear, predicting future values of $\mu$ is both more challenging and practical than using an interspersed test set; a temporal division of the test set requires $w_{ij}$ to be robust to substantial changes in overall contact distribution, which will inevitably occur in the final 18\% of a data set. In contrast, an interspersed train-test split reduces the error of our model, but some of this improvement may be attributed to learning the wear trend, not a true \textit{predictive} connection between contact distribution and $\mu$. Thus we do not report results obtained in this manner.

\begin{figure} 
\includegraphics[width = .48\textwidth]{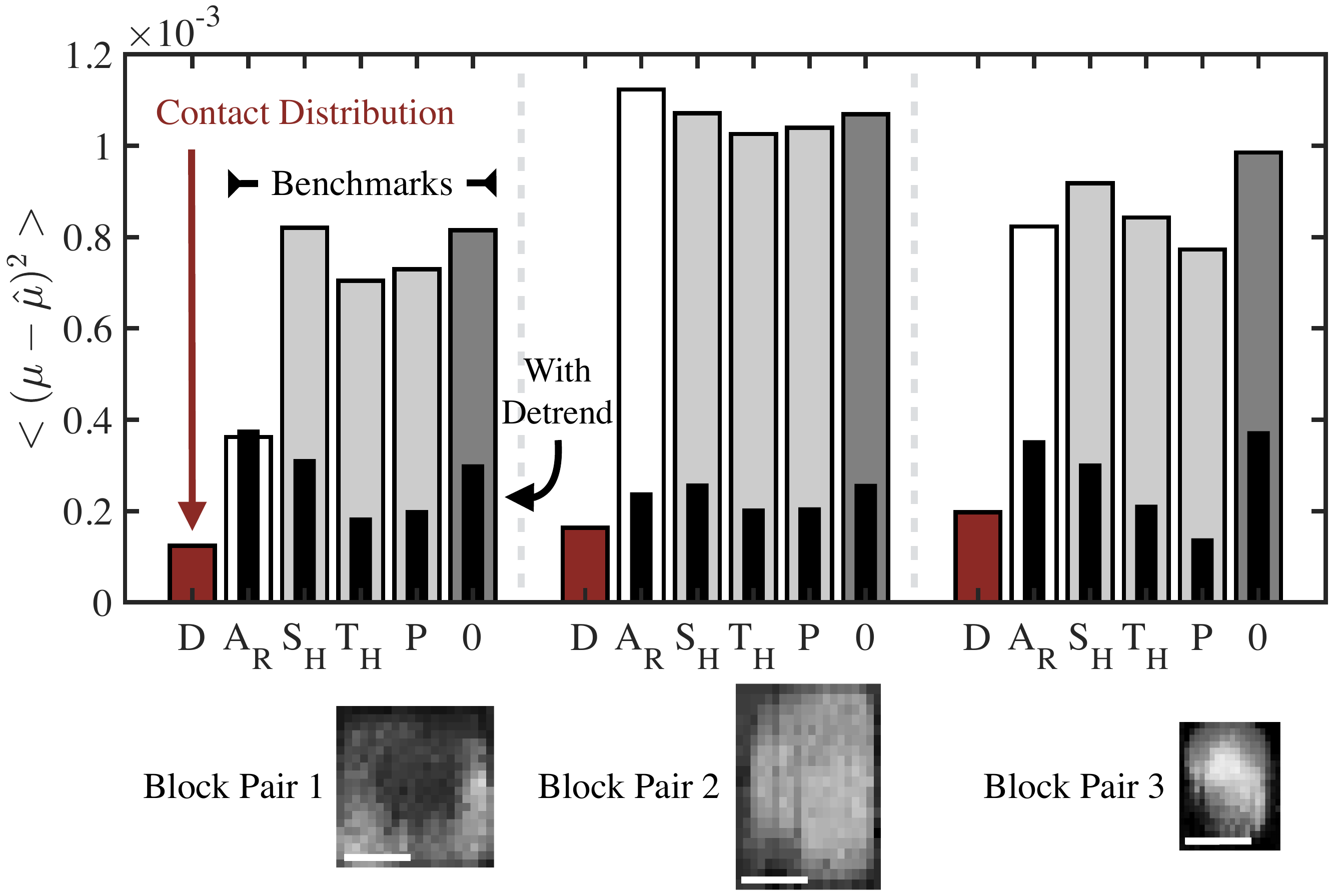}
\caption{\textbf{Contact Distribution Encodes Frictional Strength} Test-set mean squared prediction error for three distinct block pairs. Our linear regression method using the distribution of contact ($D$) is shown in red and is compared with four methods using global variables. The errors of optimal experimental predictors (Eq.~\eqref{experimental}) using hold time $T_H$, hold shear $S_H$, and both parameters ($P$) are shown in gray. Prediction error for a linear fit to the total area of contact ($A_R$) is shown in white. Error from predicting the mean value ($0$) is shown in dark gray. Thin black bars indicate error when using these benchmark methods on $\mu$ with the slow trend from wear removed as described in text. 10x10 max-pooled example images from each block pair are shown below prediction errors, to scale with one another. Scale bars are 5mm.}
\label{3}
\end{figure} 

Prediction using contact distribution performs strikingly well, as shown in Fig \ref{3}. The most obvious benchmark for comparison is prediction using a linear fit to the \textit{total} contact area $A_R$,
\begin{equation}
    \hat\mu_A(A_R) = a + b A_R \quad \quad \quad A_R = \sum_{ij} p_{ij},
\end{equation}
where $a,b$ are fitting parameters. This gives an error 3 to 7 times higher than our distribution-based predictor. It is worth emphasizing that our regression model is quite distinct from the classical method of aggregating contact area; for our system, with only a single normal load, we find variations in total $A_R$ to only weakly predict variations in $\mu$.

Another natural benchmark is the optimal predictor that has access to all experimentally-controlled parameters. That is, a predictor that predicts the mean friction coefficient conditioned on the protocol,
\begin{equation}
    \hat \mu_{exp} (s,t) = \operatornamewithlimits{mean}_{\left\{s,t\right\}}\big\{ \mu^{(n)} \big\}
    \label{experimental}
\end{equation}
where the mean is taken only over experiments with the specified $S_H$ and $T_H$. As seen in Fig.~\ref{3}, $\hat \mu_{exp}$ is only modestly better than predicting the unconditioned mean value, generating at least 4 times higher error than our methods. As previously mentioned, these parameters, along with $A_R$, \textit{do} correlate with $\mu$, however the signal is drowned in noise for individual experiments, and the relationship between these data and $\mu$ may evolve as the interface wears.

It is interesting to ask what the model is and is not learning through regression. Unfortunately, we cannot directly interpret $w_{ij}$ to `understand' the learned interfacial state: the problem is largely over-parameterized and therefore there are many different $w_{ij}$ that give similar prediction metrics. For example, using different weight regularization methods, such as LASSO or similar techniques, provides vastly different weights, with comparable predictive power. This makes the weights themselves problematic to interpret directly. Nonetheless, the regression \textit{is} learning aspects of this interfacial system that apply beyond its training set, as seen by the low error on the test set.

\begin{figure} 
\includegraphics[width = \columnwidth]{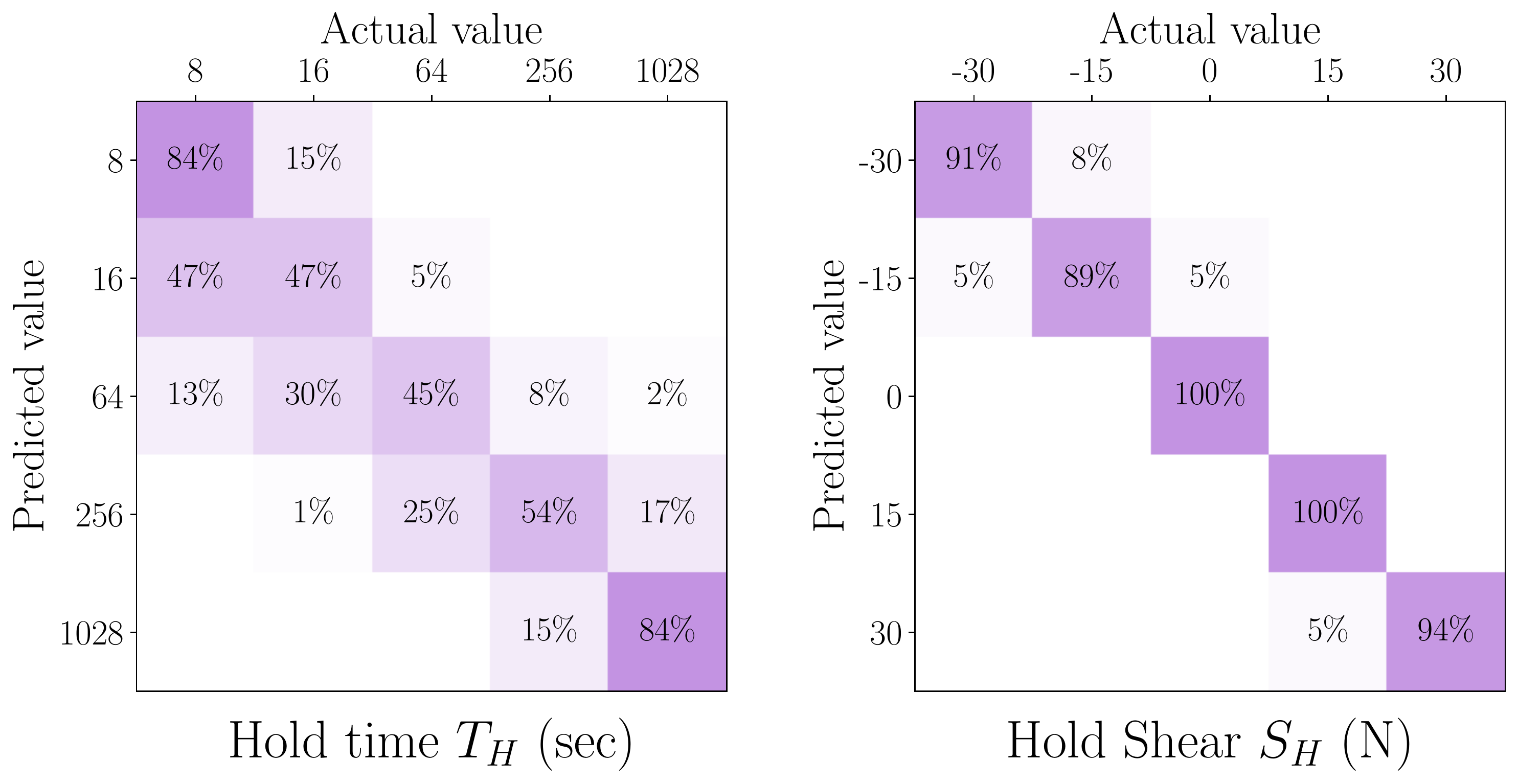}
\caption{\textbf{Experimentally controlled parameters can be extracted from the contact distribution}. The percentage of actual and predicted $S_H$ and $T_H$ for  block \#2. Other blocks show qualitatively similar results.
}
\label{4}
\end{figure} 

Since the weights are not directly interpretable, we must consider the possibility that the model is not learning anything but the connection between $mu$ and the three variables that account for much of its variance: $S_H$, $T_H$ and interfacial wear, all of which are encoded in the contact distribution. If this is so, predictions using these three factors should perform at least as well as our model. To give our benchmarks access to the evolution due to wear, we de-trend the friction coefficient by defining
\begin{equation}
    \tilde \mu^{(n)} = \mu^{(n)} - f(n)\ ,
\end{equation}
where $f$ is a low-order polynomial fit to the training data (different for each block, shown for block pair 1 in Fig \ref{1}(c) as a purple line). This is similar to methods employed in previous works to increase the signal-to-noise ratio of the evolution of $\mu$ as a function of hold time \cite{Dillavou:2018in,Dillavou:2020cs}. 

When trained to predict $\tilde\mu$ the benchmarks $\hat \mu_{exp}$ and $\hat \mu_A$ perform far better than when they are trained to predict $\mu$, as shown by the thin black bars in Fig \ref{3}. However, they are still typically worse than our distribution-based predictor \textit{while trained using the raw} $\mu$. This suggests the connection between the contact distribution and $\mu$ found by our model is not solely mediated by $S_H$, $T_H$, and wear. As further evidence, our model is only passable at predicting these experimental parameters. When trained using the same training/test split, but to predict the values of $S_H$ or $\log($ instead of $\mu$, our model predicts $S_H$ correctly $\pm$5N in 99\%, 74\%, and 95\% of cases , and $T_H$ within a factor of 2 in 62\%, 73\%, and 100\% of cases for block pairs 1, 2, and 3 respectively. The confusion matrix of block \#2 is shown in Fig.~\ref{4} as a representative example. 

If asked to predict $\mu$, a skilled experimentalist with access to all confounding variables might optimally choose $\hat \mu_{exp}$ trained on the detrended friction coefficient. Two experiments that are performed with the same protocol with relatively close $n$ are ``experimentally identical'' and all further variation beyond that connected to $S_H$ and $T_H$ is signal which is unaccounted for. Our model outperforms this predictor. Interestingly, unlike the accuracy of the benchmarks, the accuracy of our model does not improve by de-trending the data, and in some cases fairs slightly worse, depending on the de-trending protocol. This behavior is consistent with the idea that the contact distribution encodes the raw strength, and therefore predicting the actual $\mu$ values is actually easier than learning the de-trended $\tilde \mu$. To predict a de-trended value of the friction coefficient, a model must simultaneously learn a connection to $\mu$ \textit{and} the subtracted trend, which is not trivial to project forward in time. 

We have shown that the distribution of interfacial contact encodes information about frictional strength. Using a simple linear model and direct measurements of the real area of contact, we are able to predict future measurements of static friction in an experimental system. These predictions outperform more standard (averaging) predictions using $A_R$ and experimental parameters, and even typically outperform these benchmarks when the overall trend of wear is subtracted from the data. 

It is possible that regions given high weight are regions that contain weak contact or high residual stress likely to nucleate slip, or regions that contain `barriers,' strong contact regions that stop fledgling slips from propagating to the entire system \cite{Das:1977gl}. We tried several ways to tease out these details. We could not obtain discernible improvements over the linear model by using neural nets, neither terms of error nor explainability (as expected). We also trained predictors using only subsections of the interface, widely varying the size and location of these regions.  However, since these problems are so overwhelmingly over-parametrized we could not draw any consistent conclusions from the results. 

Of course, our linear regression model contains neither the as-yet unsolved mechanics of slip nucleation nor the dynamics of frictional aging, and our solutions $w_{ij}$ are not transferable or general, as they are based on the specific details of a data set from a single pair of blocks. However, our model has shown that correlating features of a map of contact points to frictional strength is feasible and out-performs traditional predictions including total quantity of contact. As we cannot measure contact quality, a direct comparison is impossible, however it is worth emphasizing that our model has no access to contact quality; pixels in our images even prior to max-pooling are on the scale of single contacts, and our model is unable to reliably predict hold time, a factor known to correlate with contact quality. Thus, when our model uses the distribution of contact to eliminate the majority of error produced when using total quantity of contact as a predictor, it implies a strong connection between contact distribution and frictional strength, as a possible alternative to the often un-measurable contact quality. Generalizing our approach to encode physical knowledge, both in the model and the regularization, is a promising avenue for future work. 

\begin{acknowledgments}
This research was funded by the National Science Foundation through DMS-1715477, MRSEC DMR-1420570, ONR N00014-17-1-3029 and the Simons Foundation. SD acknowledges funding from the Smith Family Fellowship.
\end{acknowledgments}

\bibliography{bib}

\end{document}